\documentstyle[multicol,aps,eqsecnum,psfig]{revtex}
\begin{document}

\preprint{\vbox{\hbox{U. of Iowa preprint 97-2502}}}

\title{
High-Accuracy Calculations of the Critical Exponents of Dyson's 
Hierarchical Model}

\author{J. J. Godina \\
{\it Dep. de Fis. , CINVESTAV-IPN , Ap. Post. 14-740, Mexico, D.F. 07000 \\
and
Dpt. of Physics and Astr., Univ. of Iowa, Iowa City, Iowa 52242, USA}}
\author{Y. Meurice and M. B. Oktay\\ 
{\it Dpt. of Physics and Astr., Univ. of Iowa, Iowa City, Iowa 52242, USA}}

\maketitle

\begin{abstract}
We calculate the critical exponent $\gamma $ of Dyson's hierarchical model by 
direct fits of the zero momentum two-point function 
calculated with an Ising and a Landau-Ginzburg measure,
and by linearization about
the Koch-Wittwer fixed point. We find $\gamma= 1.299140730159\ \pm 10^{-12}$.
We extract three types of subleading corrections (in other words, 
a parametrization of the way
the two-point function depends on the cutoff) from the 
fits and check the value of the first subleading exponent from the linearized 
procedure. We suggest that all the non-universal quantities
entering the subleading corrections 
can be calculated systematically from the non-linear contributions 
about the fixed point and that this 
procedure would provide an alternative way to introduce the bare
parameters in a field theory model.
\end{abstract}
\begin{multicols}{2} \global\columnwidth20.5pc
\multicolsep = 8pt plus 4pt minus 3pt

\section{Introduction}
\label{sec:intro}

Scalar field theory has many important applications in condensed matter
and particle physics. Unfortunately,
there exists no approximate treatment of this theory which could pretend
to compete in accuracy with  perturbative methods in 
quantum electrodynamics at low energy.
Accurate calculations of subtle effects at accessible energies 
provide a window on hypothetical
high energy degrees of freedom which are not
accessible by production experiments. If we imagine for a moment that
the kaons were the heaviest particles that we could produce, a precise
determination of their weak matrix elements would become a unique way
to obtain a quantitative information 
about the charmed quark. Rescaled versions of this imaginary situation 
may become relevant in the future.  

The main goal of this article is to demonstrate that the use of 
hierarchical approximations
allows determinations of the renormalized
quantities, with an accuracy which can compete with perturbative QED,
and for a wide range of UV cutoff and bare parameters.
The use of hierarchical approximations 
simplifies the renormalization group (RG) transformation while 
preserving the qualitative features of scalar field theory.
Well-known examples are 
the approximate recursion formula
derived by K. Wilson\cite{wilson}, or the related recursion formula
which holds for Dyson's hierarchical model\cite{dyson}. 
If used for quantitative purposes, hierarchical approximations need
to be improved. This is a difficult task under investigation and
which is not discussed here.
If the hierarchical approximation could be improved in a way which
maintain the advantages of the approximation,
one could obtain results with an accuracy which would 
outperform any Monte Carlo method and defy
any experimental patience. 

In the following, we use the scaling laws\cite{parisi} 
to express the renormalized 
quantities as function of the bare quantities and a UV cut-off. This 
parametrization (see e.g., Eqs. (\ref{eq:param}) and (\ref{eq:cutoff}))
is motivated below.
We show with an example that the unknown parameters entering in
the scaling laws can be determined accurately. In particular,
one universal quantity entering in the scaling 
law associated with
the two point function (the critical exponent $\gamma$) can be 
calculated with two independent methods with an agreement in the 
12-th decimal point. These two methods were 
sketched in Ref. \cite{rapid}. In the meantime, these methods were
improved in order to get a significantly better accuracy.
A detailed description of these two methods is the main technical 
content of the present paper.

All the calculations reported in this paper were made with
a specific example selected for its simplicity. However, there
is nothing essential in the choice of this example and 
accurate calculations can be performed with the same tools
in a much broader range of models and parameters.
For the sake of definiteness we now describe the example 
chosen hereafter. 
We have limited the discussion 
to a calculation of the zero-momentum two point
function of Dyson's hierarchical model with a one-component
scalar field and with a choice of the free parameter (denoted
$c$) which approximates a $D=3$ theory. 
We considered a wide range
of UV cutoff (14 orders of magnitude) 
and two different sets of bare parameters. 
The choice of the hierarchical model is not essential either.
Wilson's approximate recursion formula is closely related to the recursion
formula appearing in Dyson's hierarchical model\cite{dyson}. 
It is possible to continuously interpolate between the 
critical exponents of the two cases\cite{fam}. 
The numerical treatment of the two cases is completely
identical.
We have specialized the discussion
to the case of Dyson's model because this model
has been studied\cite{prev,wittwer,high,osc,finite} 
in great detail in the past and because a fixed point
of this model at $D=3$ is known with great precision\cite{wittwer}.
For the sake of completeness the main features of Dyson's model 
are reviewed in section \ref{subsec:dyson}.
The choice of the two-point function is not essential.
The methods can  be extended 
to other renormalized quantities as explained in Ref. \cite{finite}. 

The goal of a typical field theory calculation, 
is to obtain the renormalized quantities 
corresponding to 
the bare parameters entering in an action and a given UV cut-off
$\Lambda $.  
In the following calculations, 
the bare parameters will appear in a local measure
of the Landau-Ginzburg (LG) form:
\begin{equation}
W_0(\phi)\propto exp^{-({1\over 2}m^2 \phi^2+ g\phi^{2p})}\ .
\label{eq:lg}
\end{equation}
The UV cut-off corresponds to the scale where the theory under 
consideration stops being an accurate description and a more 
complete or more fundamental theory is required. 
Given a set of bare parameters and a UV cutoff $\Lambda$, one can try 
to integrate\cite{wilson} the degrees of freedom between $\Lambda$ and a lower
energy scale of reference $\Lambda_R$
in order to obtain an effective theory describing phenomena at
scales below $\Lambda_R$. As an example, if we are interested in
low energy processes involving pions, 
$\Lambda$ could be chosen around $m_\rho$ and $\Lambda _R$ around 
$m_\pi$. This gives a ratio of about 6 between the two scales.
Similar ratios may be applicable for an effective description
of the Higgs boson in the hypothetical case that it results from an underlying 
strongly interacting theory.
If we are interested in the effects of the charmed quarks in non-leptonic
decays of kaons, the ratio of the two scales would approximately be 3.
Larger ratios appear if, for instance, we are interested in the effects
of the top quark in systems involving only the five other quarks, or
the effects
of the $W$ and $Z$ gauge bosons on the propagation of an an electron
in a constant magnetic field. From these examples it is clear that
one would like to be able to cover a wide  range of values of $\Lambda_R /
\Lambda$.

In the scalar theory under consideration here, the cutoff
is lowered by discrete steps which reduce the initial cutoff
by a factor $2^{-{1\over D}}$ for a ``$D$-dimensional'' theory
(the notion of dimensionality is explained at length in section
\ref{subsec:dim}).
The limit of a large UV cut-off $\Lambda$ 
can be reached by fine-tuning $\beta$, the 
inverse temperature
in Dyson's formulation of the model\cite{dyson}. 
We use here the statistical mechanics language: the
magnetic susceptibility $\chi$ is studied by varying
$\beta$, keeping the bare parameters fixed.
We could have, in a completely equivalent way,
called the susceptibility the zero momentum two point
function, set
$\beta$ -- the kinetic term coupling constant -- 
equal to one and fine-tuned the bare mass.

We now follow Ref. \cite{wilson} and 
consider a sequence $L=1,2\dots$ of models with 
$\beta=(\beta _c  -\lambda_1 ^{-L} \mu)$
where 
$\lambda_1 $ is the largest eigenvalue of 
the linearized renormalization group
transformation and $\mu$ an arbitrary positive 
parameter. If we consider a fixed value of $L$ and if
$\mu$  is of order one,
it takes about $L$ iterations of the renormalization
group transformation to get an effective theory with
a mass of order one in $\Lambda_R$ units. This comes from
the fact that in the linearized approximation (see \ref{subsec:lin}), 
$\beta_c-\beta$ measures how far we are away from the stable
manifold\cite{parisi} 
and this quantity is multiplyed by $\lambda_1$
at each iteration until it reaches a value of order one and 
the linearization does not hold any more.
This suggests the definition of the
renormalized mass $m_R^2$:
\begin{equation}
m^2_R =
{\Lambda_L^2 \over {\chi(\beta _c  -\lambda_1 ^{-L} \mu)}} \ ,
\end{equation}
where 
$\Lambda_L$ is the UV cut-off 
defined as 
\begin{equation}
\Lambda_L=2^{L\over D}\Lambda_R \ .
\end{equation}
For $\beta $ close 
enough to $\beta_c$ (i.e , for $\Lambda$ large enough), 
one can approximate the 
susceptibility with
an expression which, when $D<4$, takes the form \cite{parisi}
\begin{equation}
\chi\simeq (\beta _c -\beta )^{-\gamma } (A_0 + A_1 (\beta _c -\beta)^{
\Delta }+\dots )\ ,
\label{eq:param}
\end{equation}
where $A_0,\ A_1,\dots$ are functions of the bare parameters only.
From the above equations and the expression of $\gamma$ given 
in Eq. (\ref{eq:gamma}) which implies  
\begin{equation}
\lambda^\gamma ={2^{2\over D}},
\end{equation}
one obtains
\begin{equation}
m_R^2={{\Lambda_R^2 \mu^\gamma}\over{A_0+A_1({\Lambda_R\over
\Lambda_L})^{\Delta \over{2\gamma}}+\dots}} \ .
\label{eq:cutoff}
\end{equation}

The main technical endeavor pursued in this article 
is to determine numerically the 
unknown quantities
in Eq. (\ref{eq:param}) and to determine the nature of the next
corrections.
We have used two independent methods. The first one consists in 
fitting the susceptibility at various values of $\beta$ using
Eq. (\ref{eq:param}). The second method consists in calculating 
the eigenvalues of the linearized
renormalization group transformation in order to determine
the critical exponents. The present article contains the details of the 
results announced in Ref. \cite{rapid}. In the meantime we have
refined some of the procedures used previously and improved significantly
the accuracy of our results (e.g., at least four more significant digits
in $\gamma$). These refinements are reported in the present article. 

The first estimation of the unknown 
quantities is based on a method of calculation presented in Ref. 
\cite{finite} where it is shown that the use of 
polynomial approximations in the 
Fourier transform of the recursion relation allows efficient 
and highly accurate 
calculations of the zero-momentum Green's functions in the 
symmetric phase. The method is reviewed in section \ref{sec:cal}
where we also justify the introduction of a dimensionality parameter
and review the linearization procedure.
In section \ref{sec:err}, we analyze the errors 
associated with the method. We first explain how to get rid of the 
volume effects for the range of temperature considered later. 
We then analyze the round-off errors in 
arithmetic operations and show how to reduce them to an acceptable 
level by the use higher precision arithmetic, when necessary. 
We then discuss the effects of the polynomial truncations and of 
the numerical errors in the calculation of the Fourier transform of the 
measure given by Eq. (\ref{eq:lg}) and show that they can be
reduced to a level where they will play no role in the discussion
which follows. We conclude the section
with an explanation of why a great numerical accuracy 
can only be achieved in the symmetric phase. 

In section \ref{sec:fits}, we fit the susceptibility with the four 
parameters appearing in Eq. (\ref{eq:param}), neglecting the 
next subleading corrections indicated by the $+\dots$. After
four successive refinements, all based on reproducible linear 
fits, we obtain values of $\gamma $ with a numerical stability
up to the 13-th decimal point. In section \ref{sec:sub} we analyze
the next subleading corrections and show that neglecting them
affects slightly the 12-th decimal point in
$\gamma$. All the calculations in these two sections have been
done with two different measures and gave compatible results 
for the  universal quantities ($\gamma$ and $\Delta$). 

The most efficient way to calculate the critical exponent 
$\gamma$ and $\Delta$
consists in using the linearized 
renormalization group transformation\cite{parisi} near a fixed point.
This is done explicitly in section \ref{sec:eig}. 
In ref. \cite{rapid}, we gave convincing arguments indicating the 
uniqueness of the non-trivial fixed point for a large class of theories.
In this article, we take this uniqueness for granted and we use
the very accurate expression of this fixed point obtained from
the work of Koch and Wittwer \cite{wittwer}
rather than the less accurate fixed points used in Ref. \cite{rapid}. 
This allows
us to obtain values of $\gamma$ with estimated errors of less than
one in the 13-th significant digit, the actual value agreeing with the 
previous estimate within the expected uncertainties. 

If there is only one non-trivial fixed point 
(universality) and if we can
calculate accurately the exponents, the task of calculating
the renormalized quantities for a particular set of bare parameters
reduces to the determination of quantities like $A_0$,
$A_1,\dots$. This task can be achieved by repeated subtractions 
as shown in sections \ref{sec:fits} and \ref{sec:sub}. We are convinced
that such calculations could be performed more efficiently by using 
the fixed point and a calculation of the nonlinear effects. We have
checked\cite{steve} that such a calculations can be satisfactorily 
performed in 
simplified versions of the basic recursion relation. If this task
can be successfully completed for the model considered here, this
would mean that the precise knowledge of the fixed point 
provided by Ref. \cite{wittwer} is equivalent to 
a solution of the model. 

An accurate determination of the universal exponents provides a 
new approach of the renormalization procedure: we could try
to treat as much as possible of the $A_0$,
$A_1,\dots$ and the corresponding quantities for the higher point functions
as {\it input} parameters. This of course suppose that we have 
a detailed knowledge of their relative dependences. This question
is under investigation with various methods.

\section{The Model and the Calculation of the Susceptibility}
\label{sec:cal}

In this section, we 
describe the method used to calculate the 
magnetic susceptibility and introduce some definitions which will 
be used later. For the sake of being self-contained, we first recall
basic facts about Dyson's Hierarchical Model
\cite{dyson}.

\subsection{Dyson's Model}
\label{subsec:dyson}

The model has $2^{n_{max}}$ sites. We label the sites with $n_{max}$
indices $x_{n_{max}}, ... , x_1$, each index being 0 or 1. In order to
understand this notation, one can divide the $2^{n_{max}}$ sites into
two blocks, each containing $2^{n_{max}-1}$ sites. If $x_{n_{max}}=0$,
the site is in the first box, if $x_{n_{max}} = 1$, the site is in the
second box. Repeating this procedure $n$ times (for the two boxes,
their respective two sub-boxes , etc.), we obtain an unambiguous
labeling for each of the sites. 

The non local part of the action of Dyson's Hierarchical model reads
\begin{equation}
H =
-{\beta\over2}\sum_{n=1}^{n_{max}}({c\over4})^n\sum_{x_{n_{max}},...,x_{n+1}} 
(\sum_{x_n,...,x_1}\phi_{(x_{n_{max}},...x_1)})^2 \ .
\label{eq:ham}
\end{equation}
The index $n$, referred to as the `level of interaction' hereafter,
corresponds to the interaction of the total field in blocks of size $2^n$.
The constant $c$ is a free parameter which control the decay of the 
iterations with the size of the boxes and 
can be adjusted in order to mimic a $D$-dimensional model. This
point is discussed in more detail below.

The field $\phi_{(x_{n_{max}},...,x_1)}$ is integrated over a local
measure which needs to be specified. In the following, we will work
with the Ising measure, $W_0(\phi) = \delta(\phi^2-1)$ and the
Landau-Ginsburg measure of the form given in Eq. (\ref{eq:lg}).
The hierarchical structure of Eq. (\ref{eq:ham}), allows us to integrate
the fields while keeping their sums in boxes with 2 sites.
This can be expressed through the 
recursion relation  
\begin{eqnarray}
W_{n+1}(\phi) =
{C_{n+1}\over2}e^{{(\beta/2)}({c/4})^{n+1}\phi^2} \times \nonumber \\
\int
d\phi^{'}W_n({(\phi-\phi^{'})\over2})W_n({(\phi+\phi^{'})\over2})\ ,
\end{eqnarray}
where $C_{n+1}$ is a normalization factor which can be fixed at our
convenience. 

\subsection{Polynomial Truncations}

Introducing the Fourier representation
\begin{equation}
W_n(\phi) = \int{dk\over2\pi}e^{ik\phi}\widehat{W_n}(k)\ ,
\end{equation}
and a rescaling of the source by a factor $1/s$ at each iteration,
through the redefinition
\begin{equation}
R_n(k) = \widehat{W_n}({k\over s^n}) \ ,
\end{equation}
the recursion relation becomes
\begin{equation}
R_{n+1}(k) = C_{n+1}
exp(-{1\over2}\beta({c\over4}s^2)^{n+1}{\partial^2\over\partial
  k^2})(R_n({k\over s}))^2.
\label{eq:rec}
\end{equation}
The rescaling operation commutes with iterative integrations and the
rescaling factor $s$ can be fixed at our convenience.

We will fix the normalization constant $C_n$ is such way that
$R_n(0)=1$. Then, $R_n(k)$ has a direct probabilistic interpretation. 
If we call $M_n$ the total field $\sum\phi_x$ inside blocks
of side $2^n$ and $<...>_n$ the average calculated without taking into
account the interactions of level strictly larger than $n$, we can
write
\begin{equation}
R_n(k) = \sum_{q=0}^{\infty}{(-ik)^{2q}\over 2q!}{<(M_n)^{2q}>_n\over 
s^{2qn}} \ .
\label{eq:gen}
\end{equation}
We see that the Fourier transform of the local measure after $n$
iterations generates the zero-momentum Green's functions calculated
with $2^n$ sites and can thus be used to calculate the renormalized
mass and coupling constant at zero momentum. 

In the following, we use 
finite dimensional approximations of degree
$l_{max}$ of the form:

\begin{equation}
R_n(k) = 1 + a_{n,1}k^2 + a_{n,2}k^4 + ... + a_{n,l_{max}}k^{2l_{max}}\ .
\end{equation}
After each iteration, non-zero coefficients of higher order
($a_{n+1,l_{max+1}}$ etc. ) are obtained, but not taken into account 
in the next iteration. More explicitly, the recursion 
formula for the $a_{n,m}$ reads :
\end{multicols} \global\columnwidth42.5pc
\vskip -2.8cm \begin{picture}(290,80)(80,500)\thinlines \put(
65,500){\line( 1, 0){255}}\put(320,500){\line( 0, 1){
5}}\end{picture}

\begin{equation}
a_{n+1,m} = {
{\sum_{l=m}^{l_{max}}(\sum_{p+q=l}a_{n,p}a_{n,q}){[(2l)!/(l-m)!(2m)!]}({c/4})^l[-(1/2)\beta]^{l-m}}\over{\sum_{l=0}^{l_{max}}(\sum_{p+q=l}a_{n,p}a_{n,q}){[(2l)!/l!]}{(c/4)^l}[-(1/2)\beta]^l}} \ .
\label{eq:alg}
\end{equation}
\vskip -2.7cm\begin{picture}(290,80)(80,500)\thinlines \put(
330,500){\line( 1, 0){255}}\put(330,500){\line( 0, -1){
5}}\end{picture}

\begin{multicols}{2} \global\columnwidth20.5pc

As one can see that once an initial $R_0(k)$ is given, the procedure is
purely {\it algebraic}. The 
initial conditions for the Ising measure is $R_0(k)=cos(k)$. For
the LG measure, the coefficients in the $k$-expansion need to be
evaluated numerically. 
This method has been discussed and tested at length in Ref.\cite{finite}.
The dimension $l_{max}$ of the polynomial spaces required to make reasonably 
accurate calculation is remarkably small: less than 50 
for a typical calculation (see Ref.\cite{finite}
for details).

As far as numerical calculations are concerned, 
the choice of $s$ is a matter of convenience. For the calculations in
the high temperature phase (symmetric phase) not too close to the
critical points, or for high temperature expansions the choice $s=\sqrt2$
works well \cite{high,osc}. On the other hand, the choice
of rescaling factor $s=2c^{-1/2}$ prevents the
appearance of very large numbers when we are very close to the
critical temperature. 
In the following, the finite volume magnetic susceptibility is defined as 
\begin{equation}
\chi_n (\beta) = {<(M_n)^{2}>_n \over {2^{n}}} \ .
\end{equation}
From Eq. (\ref{eq:gen}), we obtain
\begin{equation}
\chi_n = -2a_{n,1}({s^2\over2})^n \ .
\label{eq:resc}
\end{equation}

\subsection{Introducing the Dimensionality}
\label{subsec:dim}
From a conceptual point of view, the choice  $s=2c^{-1/2}$ is of particular
significance because
the
infinite volume action 
given in Eq.(\ref{eq:ham}) is invariant under the removal of the
$l=1$ terms (first level interactions) 
followed by the rescaling of the fields.
In other words, the kinetic term is not renormalized and
$\eta$ =0.
From this, we can derive
the way $c$ should be tuned in order to mimic a 
$D$-dimensional
system. Given that the dimension
of a scalar field in $D$-dimension is $[\phi] = [L]^{-{(D-2)\over2}}$  
where $L$ is a length, we obtain in the continuum
\begin{equation}
[(\int d^Dx \phi(x))^2] =L^{D+2} \  .
\label{eq:intsq}
\end{equation}
On the lattice this becomes 
\begin{equation}
[<(M_n)^2>_n] =  L^{D+2}\ . 
\label{eq:dim}
\end{equation}
If we use the rescaling factor $s=2c^{-1/2}$, the non-local part of the action
given in Eq. (\ref{eq:ham}) is invariant under a renormalization group
transformation. If in addition the local measure is also left invariant,
the average values of the even powers of the {\it rescaled} 
field stays constant. Returning to the 
original field variables, we found that at (or sufficiently close to)
a fixed point,
\begin{equation}
<(M_n)^2>_n \propto({4\over c})^n \ .
\label{eq:scaling}
\end{equation}
The only relevant scale is the size of the box
over which we have integrated all the field variables except for their 
sum. The volume of the box is proportional to the number 
of sites inside the box:
\begin{equation}
L^D\propto 2^n \ .
\end{equation}
Using this together with Eqs. (\ref{eq:dim}) and (\ref{eq:scaling})
we obtain
\begin{equation}
{4\over c}=2^{{1\over D}(D+2)}\ ,
\end{equation}
or in other words,
$c=2^{1-{2\over D}}$. 

All the calculations done hereafter have been done for $D=3$.

\subsection{Review of the Linearization Procedure}
\label{subsec:lin}
 
We now briefly review the linearization procedure.
We denote the eigenvalues of the linearized RG
transformation 
by $\lambda_n$ with the convention
$\lambda_1>1>\lambda_2>\lambda_3\dots$.
The closeness to the fixed point is essentially monitored by the 
motion along the unstable direction.
Until the number of iterations $n$ reaches a value $n^\star$ such that
$\lambda_1^{n^\star} (\beta_c-\beta) \sim 1$, $R_n$ is ``close'' to the 
fixed point and $a_{n,1}$ stays close to its fixed point value (assuming that
we use the scaling factor $s= {2\over{\sqrt{c}}}$). 
When $n$ gets larger than $n^\star$, $\chi$ starts stabilizing.
Using the relation between  $a_{n,1}$ and 
$\chi_n$ given by  Eq. (\ref{eq:resc}), we obtain the order of magnitude
estimate:
\begin{equation}
\chi\sim ({2\over c})^{n^\star} = (\beta_c-\beta)^{- {ln({2\over c})\over ln(\lambda_1)} }\ .
\end{equation}
Reexpressing in terms of $(\beta_c -\beta)$, we find that 
the exponent for the leading singularity is  
\begin{equation}
\gamma = $ln$({2\over c})/$ln$(\lambda_1)\ .
\label{eq:gamma}
\end{equation}
According to the same linear argument, the order of magnitude of the 
components in the stable directions should be proportional to 
$\lambda_l^{n^\star}$ with $l\geq 2$. Using the estimate for $n^\star$
and reexpressing in terms of $(\beta_c -\beta)$, we obtain the 
subleading exponents $\Delta_l = -$ln$(\lambda_l)/$ln$(\lambda_1)$ for
$l\geq 2$. In the following we simply use the notation
$\Delta$ for $\Delta_2$ and the higher exponents will play no significant
roles.

\section{Error Analysis}
\label{sec:err}

There are three important sources of errors which need to be considered 
when we
calculate the magnetic susceptibility: the finite volume effects, the 
round off errors and the effects of 
the finite dimensional truncation. A general discussion of these questions 
is given in
Ref.\cite{finite}. In the following, we discuss them in the particular 
cases required for the calculations of section \ref{sec:fits}.
In addition, we discuss the effects of the errors on the initial
coefficients. All the calculations done hereafter have been 
made in the symmetric phase. In the last subsection, we
explain why the present methods do not yield accurate
results in the broken symmetry phase.

\subsection{Volume Effects}

As explained in \ref{subsec:lin}, 
when calculating the susceptibility at values of $\beta$ close to and below
$\beta_c$, we spend about $-ln(\beta_c-\beta)/ln(\lambda_1)$ iterations near 
the fixed point.
During these iterations, we have the ``conformal'' scaling of 
Eq.(\ref{eq:scaling}) and the round-off errors
are amplified along the unstable direction (see next subsection). 
After that, assuming we are in the symmetric phase, 
the order of magnitude of the 
susceptibility stabilizes and the corrections get smaller by a factor 
${c\over 2}$ at each iterations. At some point, all the recorded digits
stabilize (irrespectively of the numerical errors which occurred in the 
first stage described above).
This give the estimate\cite{finite} for the number of iterations
$n(\beta, P)$ to stablize $P$ digits (in decimal notations)
\begin{eqnarray}
n(\beta,P)&=&\left(
{Dln(10)\over2ln(2)} \right) [P-\gamma
log_{10}(\beta_c-\beta)]\ .
\end{eqnarray}
For $P=16$, 
$\gamma\simeq1.3$ and $\beta_c-\beta=10^{-9}$, we obtain $n\simeq140$ and 
we need to add about 7 iterations each time we get closer to $\beta_c$ by
a factor $10^{-1}$. It is thus quite easy to get rid of the volume 
effects. In the following, we will perform calculation  $\beta <\beta_c 
-10^{-14}$ and $n_{max}=180$ will be enough
to avoid finite volume effects. 

\subsection{Numerical Errors}
\label{subsec:numerr}

From Eq.(\ref{eq:alg}), we see that the calculation of each of the 
$a_{n+1,l}$ involves a number of arithmetical operations proportional
to $l_{max}$. When we are close to the fixed point, these errors
generate small contributions in the unstable direction. These errors 
are then amplified by a factor $\lambda_1$ at each iteration until we 
move sufficiently far away from the fixed point. 
Consequently, the closer $\beta$ is to $\beta_c$, the more time is spent
near the fixed point and the larger the numerical error become.
A simple calculation\cite{finite} corresponding to this reasoning
shows that the relative errors obey the approximate law
\begin{equation}
|{{\delta \chi }\over{\chi}}|\sim {{ \delta}\over{\beta _c - \beta}} \ ,
\label{eq:numerr}
\end{equation}
where $\delta $ is 
a typical round-off error. 

\begin{figure}
\vskip15pt
\centerline{\psfig{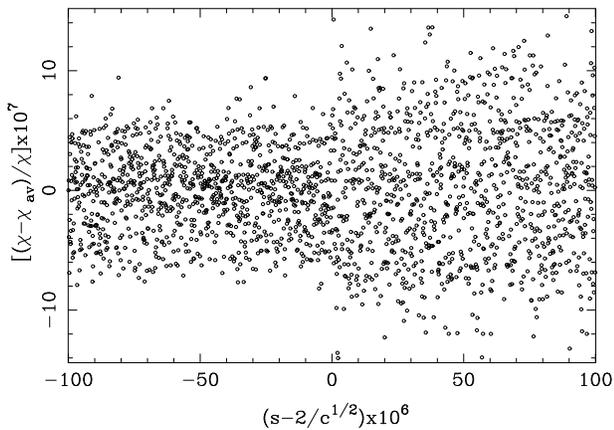}}
\vskip15pt
\caption{Distribution of the magnetic susceptibility $\chi$ with
  respect to the scaling factor $s$. }
\label{fig:spread}
\end{figure}

A simple way to probe the numerical errors is to make a small change in 
the rescaling factor $s$. 
As explained in the previous section, we can in principle use any value
of $s$ to calculate the susceptibility. This arbitrariness is compensated 
at the end by an appropriate rescaling given in 
Eq. (\ref{eq:resc}).
If we could perform the arithmetic
operations exactly, the susceptibility would be completely independent of
$s$. However, due to the round-off errors, the susceptibility actually 
depends on
$s$. 
This is illustrated in Fig. \ref{fig:spread}
where we  calculated the distribution of $\chi$ 
for values of $s$ varying
between ${2\over \sqrt{c}}-0.0001$ and
${2\over \sqrt{c}}+0.0001$ by step of $10^{-7}$
This calculation has been performed in Fortran with double 
precision variables. We have used an initial Ising measure with
$\beta=\beta_c-10^{-9}$. This distribution has a  
mean value $\mu=1.041926904\times 10^{12}$   and a variance 
$\sigma=4.9 \times 10^5$. From these quantities, we estimate that the relative
errors $|{{\delta \chi }\over{\chi}}|$ due to numerical errors should be
of order $\sigma/\mu=4.7 \times 10^{-7}$.
This is in agreement with the order of magnitude estimate of
Eq. (\ref{eq:numerr}):
using $10^{-16}$ as a typical round-off error in a double-precision 
calculation, we obtain $|{{\delta \chi }\over{\chi}}|\sim 10^{-7}$
for $\beta_c-\beta=10^{-9}$. 
A more accurate calculation performed with methods described below gives
the result $\chi=1.041926626\times 10^{12}$. We have checked that this result
was invariant under slight changes in $s$. From this, we see that
$\mu - \chi=2.8\times 10^5$ which is approximately $0.57 \sigma$.
Another information concerning the spread of the values is the difference 
between the largest and the smallest values of the distribution which
is $6.4 \sigma$ in the present case. 

There a several detailed features 
of this distribution which are not well understood. The first one is that
the distribution is not symmetric
about $s={2\over \sqrt{c}}$. 
The values of $\chi$ spread more above $s={2\over \sqrt{c}}$.
In addition, a more detailed study shows that the distribution
is not well centered and that
of values about the mean value departs more from a gaussian distribution
than expected,
given the number of ``independent trials'' (2000 in Fig. \ref{fig:spread})
made. 
In addition, increasing the statistics does not decrease $\mu - \chi$ or
increase significantly the difference 
between the largest and the smallest value.
These questions are now being investigated with low $l_{max}$ examples.

In conclusion, we have a good control on the maximal errors made 
as a consequence of the round-off errors. These seem not to exceed 10
times the order of magnitude given by Eq. (\ref{eq:numerr}). On the other
hand, we have an incomplete understanding of their distribution within
these bounds. These precludes the use of statistical methods to
obtain more accurate results and other methods need to be used.

The most efficient way to improve the accuracy of $\chi$ consists in 
using higher precision arithmetic. This can be done easily, for instance,
using the $Mathematica$ environment where one use the instruction 
$SetPrecision[\ ]$ to introduce numbers with a desired number of significant 
digits and the instruction $Precision[\ ]$ to monitor the numerical errors.
The initial precision can then be adjusted 
empirically in order to obtain a desired accuracy for $\chi$. This accuracy 
is then checked by making changes in $s$ as explained above.
A typical calculation with $l_{max}=50$, $n_{max}=200$ and a
required accuracy of 16 digits in the final result 
takes of the order of $10^3\ sec$ on a
common workstation. The same type of calculation in double-precision Fortran
takes about $0.1 \ sec$. While the high-precision program runs, we could 
thus run the double-precision $10^4$ times. If a proper understanding of
the statistical distribution of the errors was at hand (as explained above, 
this
is not the case), we could hope 
to use the $10^4$ values to reduce $|\delta \chi |$ by a factor $10^{-2}$.
In the example discussed above, we would get hope to get
errors in the 9-th significant
digit instead of the 7-th. However, with the high-precision method we 
obtained 16 correct significant digits. In the example discussed above
we obtain the accurate value $\chi=1.0419266255... \times 10^{12}$. 
The difference between this more accurate value and the mean calculated 
above with 2000 data points is $0.57\times \sigma$ and stays at the same
large value when the statistics is increased. In the following,
the high-precision method will exclusively be used.

\subsection{Determination of $l_{max}$}

As shown in Ref.\cite{finite}, the effect of the finite truncation
decays faster than exponentially with $l_{max}$ in the symmetric phase.
In general, the determination of $l_{max}$ depends on how far we are from 
criticality and the required accuracy on the value calculated. 
In the following, we will require a 13 significant digits on 
$\chi$ and  $(\beta_c-\beta)$ with $\beta <\beta_c 
-10^{-14}$. When $\beta =\beta_c -10^{-14}$, all the significant digits 
up to the 13-th decimal point of 
the quantity $(\beta_c-\beta)$ are lost since they cancel. Consequently, 
in order to get 13 significant digit in $(\beta_c-\beta)$ in the 
range considered, we will determine $\beta_c$ with an accuracy of $10^{-27}$. 
This will be the most stringent requirement to determine $l_{max}$.
As explained 
in the previous subsection, we can easily perform calculations with 
high-precision arithmetic and follow the bifurcations\cite{finite} 
in the ratios
of successive $a_{n,1}$ in order to determine $\beta_c$.
Fig. \ref{fig:phase} shows the effect of adding or subtracting
$10^{-27}$ to $\beta_c =1.179030170446269732511874097$ in the case of an 
initial Ising
measure. 
This calculation has been performed with $l_{max}=50$. 
If we use larger  values of
$l_{max}$, $\beta_c $ remains at the quoted value.
This is just a particular example. 

\begin{figure}
\vskip15pt
\centerline{\psfig{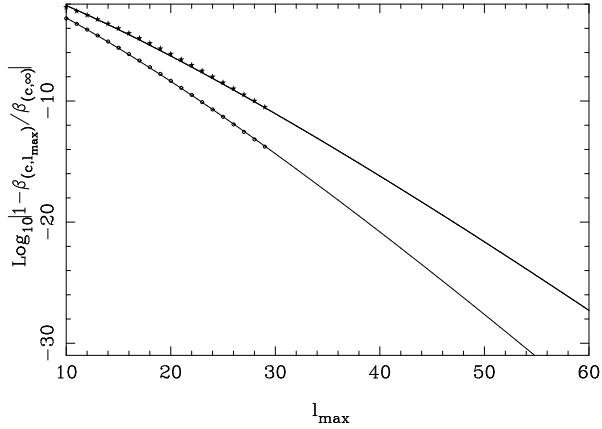}}
\vskip15pt
\caption{$log_{10}|\delta\beta_c/\beta_c|$ versus $l_{max}$
for the Ising case (circles)
and the LG case (stars). The solid line is a fit
with $a+b(l_{max}ln(l_{max}))$.}
\label{fig:delbeta}
\end{figure}

In general, the minimal 
value of $l_{max}$ for which $\beta_c $ stabilizes can be obtained from
extrapolation from the changes at low $l_{max}$ where calculations take little
time. We use the notation $\delta\beta_c$ for $\beta_c(l_{max})-\beta_c$.
The quantity
log$_{10}|\delta\beta_c/\beta_c|$ versus  $l_{max}$ 
is shown in
Fig.~\ref{fig:delbeta} for the Ising model and the LG measure of 
Eq. (\ref{eq:lg}) with $m^2=1,\ p=2,$ and $g=0.1$.
The logarithm of the relative errors falls faster than linearly.
In good approximation\cite{finite},
log$_{10}(\Delta\beta_c)\simeq a+b(l_{max}ln(l_{max}))$. So if
we want, say ${\delta\beta/\beta}\sim10^{-27}$, the choices of
$\l_{max}=48$ for the Ising case 
and $\l_{max}=60$ the LG cases appear to be a safe.
In order to check the stability of these values, 
we increased the value of $l_{max}$ to 55 in the Ising case and to 64 in 
the (LG) case, and we 
obtained the same $\beta_c$ value in both cases. We have also checked that 
these values of $l_{max}$ were sufficient to obtain 13 significant digit for 
$\chi$ in the range of $\beta$ specified above.

\subsection{Effects of the Errors on the Initial Coefficients}

\begin{figure}
\vskip15pt
\centerline{\psfig{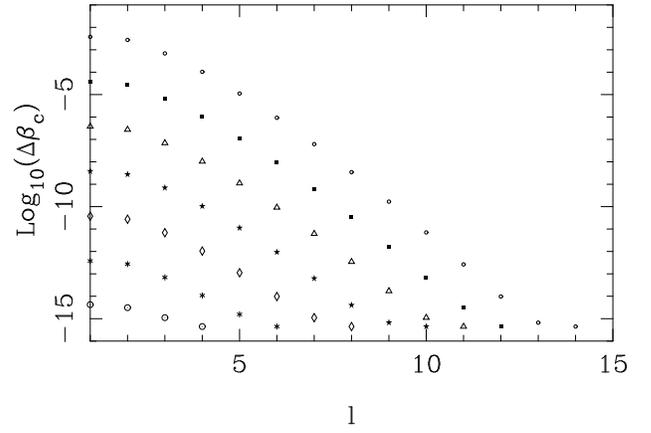}}
\vskip15pt
\caption{The shift in $\beta_c$, $\Delta\beta_c$ as a function of the relative errors in the $l$-th coefficient. ${\delta a_{0,l}\over
a_{0,l}}=10^{-2}$ (empty circles), $10^{-4}$ (filled boxes), $10^{-6}$
(empty triangles) and so on until ${\delta a_{0,l}\over a_{0,l}} =
10^{-14}$.}\label{fig:initial.ps}
\label{initial}
\end{figure}

In the Ising case, $R_0(k)=Cos(k)$, the initial coefficients are known 
analytically: $a_{0,l}={(-1)^l/(2l)!}$. However, this is not the case 
in general. 
We want to study the effect of a change in $\delta a_{0,l}$ in the
initial coefficients on $\beta_c$. 
The results are shown in Fig.~\ref{fig:initial.ps}.
The results can be read as follows. If we are interested in determining
$\beta_c$ with, say, 10 significant digits, $a_{0,10}$ has to be 
determined with 2 significant digits, $a_{0,9}$ with 3 significants digits,
$a_{0,8}$ with 4 significant digits etc... . 
In the following, we are interested in universal properties (features which
are independent of the measure) rather than in properties of particular
measures. Consequently we have only used a double-precision calculation
of the Fourier transform for the LG model. The reproducibility
of the details of the calculations then require having the same $a_{0,l}$.
On the other hand, in the Ising case, the analytical form of the initial
coefficients allows a completely reproducible procedure.

\subsection{What Happens in the Broken Symmetry Phase}

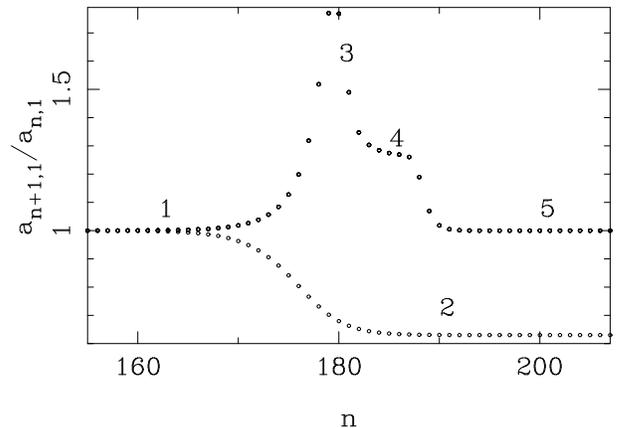
\begin{figure}
\vskip15pt
\centerline{\psfig{figure=phase.ps,height=2.2in,angle=270}}
\vskip15pt
\caption{$a_{n+1,1}/a_{n,1}$ versus $n$ for $\beta=\beta_c\mp
  10^{-27}$. }\label{fig:phase}
\end{figure}

Fig. \ref{fig:phase} shows the existence of two phases.
There are five parts of the graph we would like to discuss
here.
For the rest of the discussion, it is important to specify that 
$a_{n,l}$ has been calculated with the canonical value of the rescaling
parameter 
$s={2\over\sqrt{c}}$.
The first part is before the bifurcation. This is shown as the region
``1'' in Fig.~\ref{fig:phase} where the ratios of
$a_{n+1,1}/a_{n,1}$ are close to 1. 
The second part (``2'') shows the bifurcation in the high-temperature
phase. If we
are below $\beta_c$, the ratio $a_{n+1,1}/a_{n,1}$ will go down to the value 
$c/2$, which guarantees the 
existence of a thermodynamical limit for $\chi$ 
(since we need to multiply $a_{n,1}$ by $({2\over c})^n$
in order to get $\chi$ see Eq.(\ref{eq:resc})). 
On the other
hand, the bifurcation toward the 
low temperature phase is characterized by
a peak shown by ``3'' in Fig.~\ref{fig:phase}. 
Part 4 of the graph is a narrow
``shoulder''. In the low-temperature phase, we 
expect $<M^2_n>_n\propto 2^{2n}$, which means 
$a_{n+1,1}/a_{n,1}\simeq c\simeq 1.26$. 
We studied the
$l_{max}$ dependence of this shoulder and observed that the number of
points on the shoulder increases by approximately one when we increase 
$l_{max}$ by 10. Unfortunately, the shoulder is not infinite and after
a few iterations,
the ratios will reach to 1
again (part ``5''). This signals an attractive fixed point.
However, this is not a fixed point of the exact (not truncated)
recursion relation.
This can be seen by
looking at the coefficients $a_l^{\star}$ of these attractive fixed points
for different values of $l_{max}$. When $l_{max}$ increase, the 
values of $a_l^{\star}$ increase like $(l_{max})^l$, showing that their 
existence is due to the truncation process.

The fact that the truncation procedure generates numerical instabilities 
in the low-temperature phase can be understood from the basic 
formula Eq. (\ref{eq:rec}). In the low temperature phase, the 
measure $W(\phi)$ has two peaks symmetrically located with respect to the 
origin. At each iteration, the separation between the peaks increases by 
a factor 2 (in unrescaled units). By taking simple examples and 
going to Fourier transform, one sees that at some point the partial
sums (truncated at $l_{max}$) representing the 
exponential in Eq. (\ref{eq:rec}) becomes inaccurate because the 
argument of the exponential is too large.

\section{Critical Exponents from Fits}
\label{sec:fits}

In this section, we explain how to calculate the exponents
$\gamma$ and $\Delta$ using a sequence of 
increasingly accurate fits of the susceptibility.
The general method has been briefly outlined in Ref.\cite{rapid}.
Here, we give all the details of a significantly more accurate
calculation which leads to a determination of   
$\gamma$ with 12 decimal points.
The main ingredient of the procedure is that 
for $\beta$ close
enough to $\beta_c$ (i.e, for a cut-off $\Lambda$ large enough), one can
approximate very well the magnetic susceptibility (zero-momentum two
point function) with an expression taking into account only
the first irrelevant direction, namely :
\begin{eqnarray}
\chi &\simeq& (\beta_c-\beta)^{-\gamma}(A_0 +
A_1(\beta_c-\beta)^{\Delta}) \ .
\label{eq:sublead}
\end{eqnarray}
The estimation of the unknown quantities in this equation proceeds in
four steps.
In the first step, we
get a rough estimate for $\gamma$ by using a linear fit 
in a range of $\beta$ where we minimize the combined effects of 
the numerical errors and of the subleading corrections. 
In the second step, we 
discuss how to improve this result by estimating the sub-leading
exponent $\Delta$ and the coefficient $A_1/A_0$.
Using these preliminary estimates, we will as the third step
of the procedure, use  a
``bootstrap'' technique 
between a set of high-precision data close to criticality
and another set of data where the subleading corrections are important.
Finally, we do a linear analysis of the difference between the fit
and the high-precision data in order to get results which
are as independent as possible of the slightly arbitrary choices (how
to divide the data into ``bins'' etc... ) made during the first three 
steps. After this fourth step, we analyze the difference between the fit
and the data away from criticality and discuss the next subleading corrections.

All the calculation of these sections have been done with either an 
Ising initial measure or a LG measure of the form given in 
Eq. (\ref{eq:lg}) with $m^2=1,\ p=2,$ and $g=0.1$. We refer to these 
choices as the ``Ising case'' or the ``LG case'' hereafter.

\subsection{Localized Linear Fits}

\begin{figure}
\vskip15pt
\centerline{\psfig{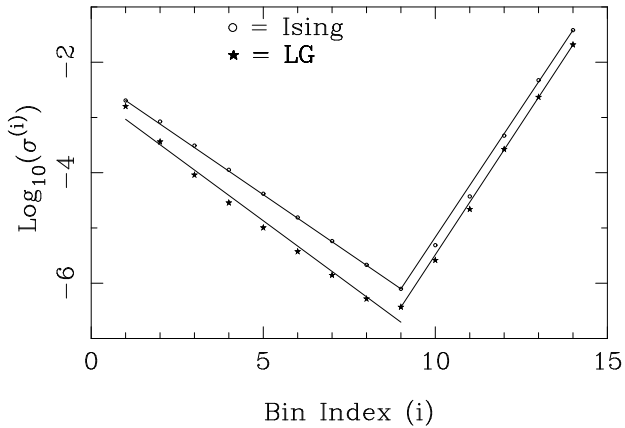}}
\vskip15pt
\caption{The deviations from the linear fits
log$_{10}\sigma ^{(i)}$ defined in the text as function of the bin index $i$,
for the Ising model (circles) and the Landau-Ginsburg model
(stars).}\label{fig:sigma.ps}
\label{sig}
\end{figure}

In the first step,
we calculate $\chi$ at various temperatures and display 
log$_{10}(\chi)$ versus -log$_{10}(\beta_c-\beta)$. We will use the notation
\begin{equation}
x\equiv -log_{10}(\beta_c-\beta)\ . 
\end{equation}
If we display log$_{10}(\chi)$
versus $x$ , we see a linear behavior with
a slope $\gamma\simeq
1.30$.
The
deviations from the linear behavior are not visible to the naked eye. 
We need to study
these deviations  locally in $\beta$. 
In order to understand the corrections,
we have divided the data into 14 bins of 100 points. 
The first bin contains the
data  $ x=1.00, 1.01, ... , 1.99 $ and so on. 
In each bin, indexed $i$ , we make a
linear fit of log$_{10}(\chi)$ versus $x$. In the $i$-th bin  
we will call the slope
$\gamma^{(i)}$ and $(\sigma^{(i)})^2$ denoted the sum of the squares of
difference between the data and the linear 
fit divided by the number of points in a
bin minus 2 which is ( for the $i$-th bin ),
\begin{equation}
(\sigma^{(i)})^2 ={
\sum_{j=1}^{100}(log_{10}(\chi_{i,j}^{data})-
log_{10}(\chi_{i,j}^{fit}))^2\over{98}} \ ,
\end{equation}
where $j$ indexes the data points in the $i$-th bin.
The values of log$_{10}(\sigma^{(i)})$ are plotted in
Fig.~\ref{fig:sigma.ps}.
It is easy to interpret this graph. There are
two major sources of deviations from the linear behavior. The first
one is the existence of subleading
corrections to the  scaling laws which decrease when $\beta$ gets
closer to $\beta_c$. As a first guess, we use
$\sigma^{(j)}\propto 10^{-\Delta j}$ so that 
log$_{10}\sigma^{(j)}\simeq -\Delta j+
constant$. By calculating the slopes between bin 1 and bin 9, we
obtain $\Delta\simeq0.42$ and $\Delta\simeq0.45$ for the Ising
and the LG cases, respectively. Thus, we already obtained a numerical
value for the subleading exponent which is roughly the same 
for the two models considered here.
The other source of deviation from the linear behavior 
comes from the numerical errors discussed in the previous section 
and which increase when
$\beta$ gets closer to $\beta_c$ according to Eq. (\ref{eq:numerr}).
$\sigma\sim{\delta/(\beta_c-\beta)}$. The slopes between bin 9
and bin 14 are $-0.95$ and $-0.96$ for the Ising and the LG cases,
respectively, in good agreement with the $ (\beta_c-\beta)^{-1}$
dependence of the numerical errors predicted by Eq. (\ref{eq:numerr}).

In bin 9, these two deviations from linear behavior 
are minimized and we can
consider $\gamma^{(9)}$ as a first estimate of $\gamma$. 
Its numerical values is 1.29917 for
the Ising case and 1.29914 for the LG case. 
By using this simple procedure we already
gained almost two significant digits compared to the existing 
estimated \cite{prev,osc} where the answer
$\gamma=1.300$ was obtained with errors of order 1 
in the last digit.

\subsection{Subleading Corrections}

The second step consist in correcting the previous estimate
by taking into account the 
subleading corrections. 
We will use the bins 6 and 7 where the next
subleading 
corrections are reasonably small and the numerical errors are not too large. 
We have divided
these two bins into 10 sub-bins of 100 points. 
We will use two digit indices for these
sub-bins. For instance, sub-bin 6.5 is the fifth sub-bin of bin 6 and 
contains the values of
$x$: 6.5, 6.501, ... ,6.599. Using the notation $\bar{x}$ for the middle of the
sub-bin and Eq. (\ref{eq:sublead}) we obtain,
\begin{eqnarray}
log_{10}(\chi(\bar{x})) & = & \gamma\bar{x} + log_{10}(A_0+A_1
10^{-\Delta\bar{x}})\ .
\end{eqnarray}
For a small change  $\delta x$ with
respect to $\bar{x}$, we obtain that at first order in this change
\begin{eqnarray}
log_{10}(\chi(\bar{x}+\delta x)) - log_{10}(\chi(\bar{x})) & = & \nonumber \\
(\gamma
-{A_1\over A_0}\Delta10^{-\Delta\bar{x}})\delta x + \vartheta(\delta x^2) \ .
\end{eqnarray}
The coefficient of $\delta x$ can be interpreted as the local slope near 
$\bar{x}$.
Indexing each sub-bin by $j$ (e.g $j: 6, 6.1, 6.2 ...$) and its
middle by
$j+0.0495$ (e.g. $6.1495$ is the middle of the sub-bin $6.1$), the 
the slope $\gamma^{(j)}$ in the sub-bin $j$ reads
\begin{eqnarray}
\gamma^{(j)} \simeq \gamma - \Delta ({A_1\over A_0}) 10^{-\Delta(j+0.0495)}
\ .
\label{eq:localg}
\end{eqnarray}

\begin{figure}
\vskip15pt
\centerline{\psfig{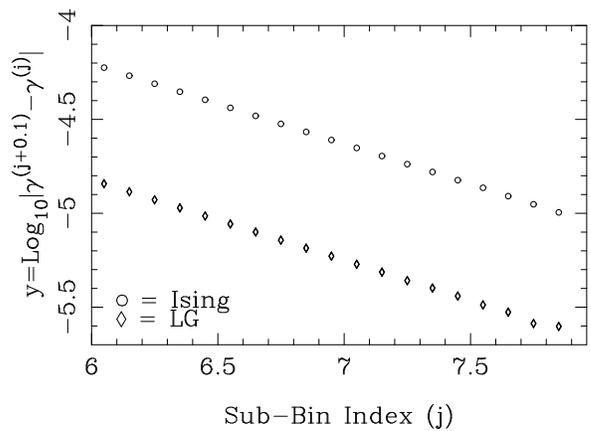}}
\vskip15pt
\caption{The linear fits of log$_{10}(|\gamma^{(j+0.1)}-\gamma^{(j)}|)$, 
for the Ising case
(circles) and for the LG case (stars).} \label{fig:step2.ps}
\end{figure}

The unknown quantities ${A_1\over A_0}$ and $\Delta$ can be obtained from  
linear fits of
log$_{10}(|\gamma^{(j+0.1)}-\gamma^{(j)}|)$  versus $j$.
From Fig.~\ref{fig:step2.ps}, one can see that in good approximation, there 
exists an approximate relationship between these two quantities in the 
two cases considered. In addition, the slope appears to be identical
for the two cases.
The slope and the intercept can be calculated from 
Eq. (\ref{eq:localg}) which implies that  
\begin{eqnarray}
log_{10}(|\gamma^{(j+0.1)}-\gamma^{(j)}|) \simeq 
 -\Delta (j+0.0495) + \nonumber \\
log_{10}(\Delta |{A_1\over A_0}|)+log_{10}(1-10^{-0.1 \Delta })\ .
\end{eqnarray}
The subleading exponent $\Delta$ is the absolute value of the slope.
Having determined $\Delta$ and knowing the intercept we can then determine
$log_{10}(|{A_1\over A_0}|)$.
Using this procedure,
we obtained
${A_1/ A_0}=-0.57$  and $\Delta=0.428$ for the Ising 
case and ${A_1/A_0}=0.14$ and
$\Delta=0.427$ for the LG case.  

If we now repeat the first step - 
a linear fit in bin 9 - but with $\chi$
divided by $[1+(A_1/A_0)(\beta_c-\beta)^\Delta]$, we obtain $\gamma=1.299141\mp
10^{-6}$ for the two models considered above. Given that for a
calculation using double precision in bin 9 we have values of $\chi$ with 
between 6 and 7 significant digits, this estimate seems to be 
the best result 
we can obtain
with this procedure (see the discussion of the numerical
errors in section III.B). 

\subsection{A Bootstrap Procedure Involving Higher Precision Data}
\label{subsec:boot}

Up to now, our desire to minimize the subleading corrections
which decay like $10^{-\Delta x}$ has been contradicted by the 
appearance of numerical errors growing like $10^{(-16+x)}$. However,
we have explained in section III.B, that it is possible to 
circumvent this difficulty by using an arithmetic having a better
precision than the usual double precision. In this subsection, 
we will 
use data having at least 13 correct significant digits 
in bin 11, 12 and 13. We call this data ``high
precision data''. As we explained before, we choose $l_{max}=50$ and
$l_{max}=58$ for the Ising and the LG case, respectively. 

Since the calculations are more lengthy, we used only 10 points in each bin. 
We also determined $\beta_c$
with 27 significant digit so that in bin 13, 
the subtracted quantity $(\beta_c-\beta)$ is also
know at least with 13 significant digit. We found $\beta_c =
1.179030170446269732511874097$ for the Ising case and
$\beta_c=1.14352915687979895500964720$ for the LG case.
We then use bin 13 (where the subleading corrections are
small and the errors are not very important) to calculate $\chi$ divided 
by the subleading
correction as explained in the previous subsection to estimate $\gamma$. 
Then with the new value of $\gamma$ obtained
we go back to bin 7 to calculate the subleading corrections.  
This procedure can be
iterated and this `bootstrap' of linear fits
converges rapidly. We obtain $\gamma=1.299140732$, $\Delta=0.4262$ and
${A_1/A_0} = -0.564$
for the Ising model and $\gamma=1.299140730$,$\Delta=0.4258$ and
$A_1/A_0=0.135$ for the LG case. These numbers change typically by one
in the last digit quoted above if one replaces bin 7 by bin 6 to evaluate 
$\Delta$. In order to remove this arbitrariness, we will now use these 
numbers as the initial values for a more accurate procedure.

\subsection{Linear Analysis of the Discrepancies}

\begin{figure}
\vskip15pt
\centerline{\psfig{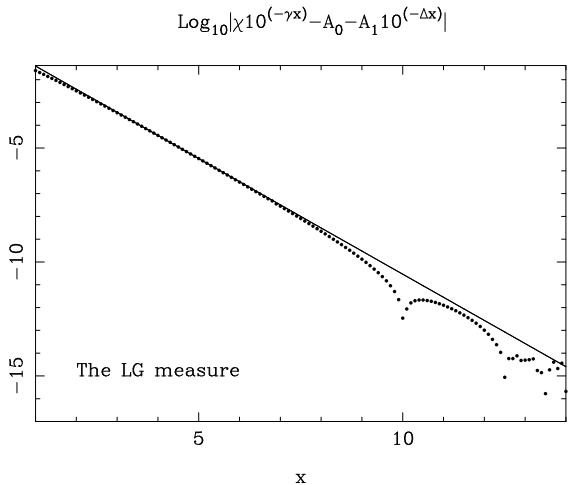}}
\vskip15pt
\caption{log$_{10}[|\chi10^{(-\gamma x)}-A_0-A_110^{(-\Delta x)}|]$ \\
  versus $x=-log_{10}(\beta_c-\beta)$ for the LG measure.
} \label{fig:subsub2}
\end{figure}

We have now reduced the errors made in the estimate of the unknown 
quantities appearing in Eq. (\ref{eq:sublead}) to a sufficiently
low level to allow us to treat these errors in a linear 
approximation. We start with an initial fit of the data, for instance
as obtained in step 3,  with
errors in  the unknown quantities parametrized in the following way:
\begin{eqnarray}
[log_{10}\chi]_{fit}=log_{10}A_0+\delta(log_{10}A_0) +
(\gamma+\delta\gamma)x + \nonumber \\ 
log_{10}(1+(c_1+\delta c_1)10^{-(\Delta+\delta\Delta)x})\ .
\end{eqnarray}
where $c_1 \equiv {A_1/A_0}$ and $\gamma$, $\Delta$ stand for the 
exact values. On the other hand, we assume that the data can be fitted
according to Eq. (\ref{eq:sublead})
\begin{equation}
[log_{10}\chi]_{data}=(log_{10}A_0+\gamma x+log_{10}(1+c_110^{-\Delta
  x})) \ .
\end{equation}
Combining the two above equations we obtain at first order
\begin{eqnarray}
[log_{10}\chi]_{fit}-[log_{10}\chi]_{data}
\simeq\delta(log_{10}A_0)+\delta\gamma x + \nonumber \\ 
 {\delta c_1 10^{-\Delta
    x}\over ln(10)} -c_1 10^{-\Delta x}x\delta\Delta \ .
\label{eq:lin}
\end{eqnarray}
Interestingly, the $x$-dependences of the four terms are all distinct
and we can fit $\delta$(log$_{10}A_0$), $\delta\gamma$, $\delta c_1$ and
$\delta\Delta$ using a standard least square procedure where the function
to be fitted depends {\it linearly} on the fitted parameters. 
This procedure can be repeated until some numerical stability is achieved.
The final results are insensitive to small changes in the initial values
coming from the uncertainties associated with the previous step.
Using bin 13, we obtain 
$\gamma=1.2991407301599$ and $\Delta=0.4259492$ for the Ising model, 
and, $\gamma=1.2991407301582$ and $\Delta=0.4259478$ for the LG case.
The small numerical fluctuations which persist
after many iterations produce changes of less than 2 in the last quoted digit.
The origin of these small fluctuations can be inferred by plotting 
$[log_{10}\chi]_{fit}-[log_{10}\chi]_{data}$ for the final fit 
(see Figs. \ref{fig:subsub1} and \ref{fig:subsub2} in the next section). 
The non-smoothness of these differences in bin 13 indicates that they 
are due to the numerical errors on $\chi$.
The amplitude
of these differences is smaller than $10^{-13}$ consistently with the 
fact that we performed the calculation of $\chi$ in a way that guaranteed 
at least 13 accurate significant digits. These fluctuations are indicative
of the limitation in the numerical precision of our procedure. The accuracy
of the value of the exponents, i.e. how close they are to the ``true'' values,
is further limited by the fact that there exist corrections to our 
main assumption Eq. (\ref{eq:sublead}). If we assume universality, 
the discrepancy between the values of the exponents for the two
cases considered
should give us an indication concerning the accuracy of the results.
For instance, the discrepancy between the two estimates of $\gamma$ is of order
$10^{-12}$ which is about ten times larger than the fluctuations of 
numerical origin. The estimation of the next sub-leading corrections
is the main topic of the next section.

\section{The Next Sub-Leading Corrections}
\label{sec:sub}

In the previous section, we have used the parametrization of 
Eq. (\ref{eq:sublead}) for the susceptibility near criticality. 
This parametrization is by no means exact and corrections 
become more sizable as we move away from criticality. 
The corrections come from effects which can be calculated by a 
linearization procedure (the next irrelevant directions, see subsection 
\ref{subsec:lin}) 
or effects
which are intrinsically non-linear. 
Anticipating the results which will be 
presented in the next section, we obtain the approximate values
$\Delta_2\simeq 0.43$ (we recall that since we only took into account
one irrelevant direction, we used the notation $\Delta$ for $\Delta_2$ 
before) and 
$\Delta_3\simeq 2.1$.  In other words if we
consider the first irrelevant direction as ``first order'', the next
irrelevant directions produce effects which are smaller than the fourth
order. In bin 13, these effects are completely
unnoticeable in our analysis. The non-linear effects are discussed in
the subsection \ref{subsec:nonlin}.
The main result obtained there is that 
all the corrections can be parametrized in the following way:
\begin{eqnarray}
\chi &\simeq& (\beta_c-\beta)^{-\gamma}(A_0 +
A_1(\beta_c-\beta)^{\Delta}+  
A_2(\beta_c-\beta)^{2\Delta} \nonumber \\
& + &  A_a(\beta_c-\beta)+\dots   )\ ,
\label{eq:nextsublead}
\end{eqnarray}
In the next  subsection, we analyze the data in terms of the new 
parametrization and extrapolate our results in order to estimate 
the errors made in the calculations of $\gamma$ and $\Delta$
in the previous section.

\subsection{Nonlinear Corrections}
\label{subsec:nonlin}
The previous analysis, describes the linearized flows near the fixed 
point. The closer to criticality we are, the more iterations are spent
close to the fixed point and the more accurate the linear description is.
Nevertheless, when we approach or leave the fixed point, nonlinear effects
are unavoidable. These nonlinear effects can be studied more easily in 
low-dimensional maps. Without entering into the detail of this 
analysis\cite{steve}, we can envision three types of corrections which we now
proceed to discuss.

As already noticed in \cite{osc}, the  ``constants'' $A_0$ and $ A_1$, should
be replaced  by functions $A_i((\beta_c -\beta)$ such that 
$A_i(\lambda_1(\beta_c -\beta))=A_i(\beta_c -\beta)$. This invariance implies
an expansion in integral powers of $(\beta_c -\beta)^{i\omega}$ 
(Fourier modes) with
$\omega={2\pi\over{ln(\lambda_1)}}\simeq 17.8$. We argue here that the 
coefficients of the non-zero powers are suppressed by 14 orders of magnitude.
The rationale for this suppression is that the non-zero modes contribute
to the extrapolated slope (an asymptotic estimator for $\gamma -1$ used with
the high-temperature expansion) with
about the same strength as the zero mode\cite{osc}. However, this is
the result of a double amplification for the non-zero modes. 
This results from the equations (3.7) to (3.10) of Ref. \cite{osc}.
First, 
when calculating the coefficients of high temperature expansion one gets
an amplification factor of order 
$|\Gamma(\gamma+i\omega)|\simeq 5\times 10^{10}$. Second, while calculating
the extrapolated slope, one gets another amplification by a factor 
$\omega^3\simeq 5\times 10^3$. Putting these two factor together, we 
obtain the claimed 14 orders of magnitude. Such a small effect is smaller 
than our numerical resolution.

Second, the singularity $(\beta_c-\beta)^{-\gamma}$ should be replaced by
$((\beta_c-\beta)+d_2(\beta_c-\beta)^2+\dots)^{-\gamma}$ with coefficients
$d_l$ calculable in low dimensional maps. These corrections generate
{\it analytical} corrections to the scaling law in contrast to the 
subleading corrections which are in general not integer powers 


Third the nonlinear corrections associated with the irrelevant
directions generate corrections which are presumably of the form
$(\beta_c-\beta)^{l\Delta}$ with $l=2,3,\dots$. Later we call
these corrections the quadratic corrections or the second order effects.

In summary, the corrections associated with nonlinear contributions
obey the parametrization of Eq. (\ref{eq:nextsublead}) for a 
sequence of exponents
$0.43,\ 0.86,\ 1, \ 1.29,\ 1.72,\ 2,\dots$. Note that these exponents
are very close to each other and it may be difficult to disentangle their 
effects.

\subsection{Empirical Determination of the Corrections}

We are now ready to use the data to determine some of the unknown quantities
in Eq. (\ref{eq:nextsublead}).
In the following, we will study these corrections
for the Ising and the LG cases separately. The reason for doing
this is that in the Ising case, the ratio
${A_1/A_0}=-0.56$ while in the LG case, ${A_1/A_0}=0.14$. 
The relative size of the quadratic corrections is presumably of order
$(A_1/A_0)^2$ and these corrections will be more
sizable in the Ising case. We start with the assumption
that there is one next subleading correction which dominates when we move
from bin 13 to smaller values of $x$.
In other words,
\begin{equation}
\chi10^{-\gamma x}-A_0-A_1 10^{(-\Delta x)}\simeq A 10^{-\phi}\ ,
\label{eq:onenext}
\end{equation}
in an intermediate $x$ region.
In this equation, 
the four parameters $\gamma$, $\Delta$, $A_0$ and
$A_1$ are understood as their best estimates near criticality obtained in
the previous section. 
Anticipating the results obtained below, the exponent $\phi$ is roughly of 
order one. The corrections in bin 8 are thus of order $10^{-8}$ which is 
precisely of the same order as the numerical errors if we use 
double-precision calculations. Consequently we had to use the
$Mathematica$-based method described 
in the section \ref{subsec:numerr} 
in order to get at least 4 significant digits
for the corrections. For time considerations, we have limited our 
calculations to 10 points per bin.

\begin{figure}
\vskip15pt
\centerline{\psfig{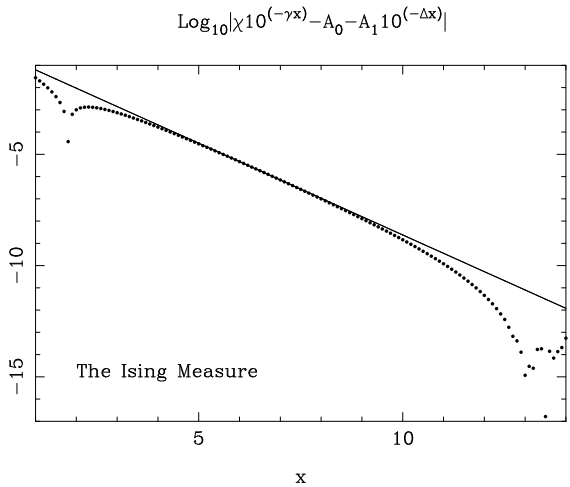}}
\vskip15pt
\caption{log$_{10}[|\chi10^{(-\gamma x)}-A_0-A_110^{(-\Delta x)}|]$ \\
  versus $x=-log_{10}(\beta_c-\beta)$ for the Ising measure.
} \label{fig:subsub1}
\end{figure}

If Eq. (\ref{eq:onenext}) is approximately correct, 
the logarithm of the l.h.s. should be approximately linear in some region
of $x$. This quantity is displayed in Fig. \ref{fig:subsub1} for 
the Ising model
and in Fig. \ref{fig:subsub1}  for the LG model.

One sees that in both case the graph is approximately linear over a 
large region of $x$. Using a linear fit in each of the bins of these regions
we obtain $\phi=0.82$ and $A=0.4$ with $\sigma=9\times 10^{-4}$ in 
bin 6 for the Ising model
and $\phi=1.01$ and $A=0.4$ with $\sigma=6\times 10^{-4}$ in 
bin 4 for the LG model.
Other bins have larger values of $\sigma$ and values of $\phi$ which change 
by a few percent while moving from bin to bin. 

In the Ising case, we have
$|{A\over A_0}|\approx ({A_1\over A_0})^2$ and $\phi\simeq 2\Delta$ and 
we interpret this as a second order (or quadratic) effect 
associated with the first irrelevant corrections. In other words, 
$A\simeq A_2$  if we follow the notations of
Eq.(\ref{eq:nextsublead}) . 
In the LG case $({A_1\over A_0})^2\simeq 0.02$ is very small 
and the dominant effect in the linear
region is the analytic correction ($\phi=1$) behavior.
In other words, 
$A\simeq A_a$ used if we follow again the notations of  
Eq. (\ref{eq:nextsublead})  . 

The departure from linearity occurs in its most extreme way 
as dips located near $x=2$ in the Ising case and $x=10$ in 
the LG case. 
These dips signal the existence of effects of opposite signs.
A plausible interpretation of the location of these dips 
is that they occur at values of $x$ where the $10^{-2\Delta x}$ prevail
over the $10^{-x}$ analytical corrections.
A detailed analysis confirms this view
for the Ising model, which allows us to
neglect the analytical corrections in bin 13. For the LG model, two
effects compete in bin 10 which is dangerously close to bin 13 where 
the parameters are fined-tuned (see below) and we were unable to get
a clear linear behavior after one more subtraction. Our most plausible
explanation is the following for the LG model. Near $x=10$, we have
$A_a 10^{-x}\simeq -A_2 10^{-2\Delta x}$ which implies $A_2\simeq 0.016$.
With this rough estimates $|{A_2\over A_0}|\approx ({A_1\over A_0})^2$ 
which is consistent with a second order effect. So if this 
interpretation is correct the quadratic effects are about twice the size 
of the analytic corrections in bin 13 for the LG model. 

In summary, we will use the assumption that in bin 13, the corrections 
are mostly second order effects and we will neglect the analytical
corrections. This assumption is well-obeyed in the Ising case
but is just an order magnitude estimate in the LG case.

\subsection{Accuracy of the Previous Estimate}

We are now in position to estimate the effects of the next subleading
corrections in the calculation of the critical exponents 
reported in the previous section. First of all, we notice that by extrapolating
the dominant linear behavior described in the previous subsection 
to bin 13, we obtain effects smaller than $10^{-11}$ in the Ising case and 
smaller than $10^{-13}$ in the LG case. This justifies treating them 
as small perturbations in bin 13. When we fitted the data in bin 13 
without taking these small effects into account, we made small
adjustments in the fitted parameters which allowed us to fit the 
data with a precision comparable to the numerical precision.
In order to get a rough estimate of how much the next 
subleading corrections led us to misestimate the exponents,
we can linearize the next subleading corrections about $x=13$.
We obtain a change of the ``apparent'' slope:
\begin{equation}
|\delta\gamma|\approx|{{A}\over{A_0}}10^{-13\phi}\phi|\ .
\end{equation}
The order of magnitude of the corresponding errors on $\Delta$
can be estimated by equating the term linear in $\delta\gamma$ 
with the term linear in  $\delta\Delta$ in Eq. (\ref{eq:lin}).
This yields:
\begin{equation}
|\delta\Delta|\approx|{{A_0}\over{A_1}}10^{-13\Delta}\delta\gamma|\  .
\end{equation}
We insist that this is only an order of magnitude estimate.
Plugging numerical values, we obtain $|\delta\gamma|=3\times 10^{-12}$
and $|\delta\Delta|=2\times 10^{-6}$ in the Ising case, and,
$|\delta\gamma|=2\times 10^{-13}$
and $|\delta\Delta|=6\times 10^{-7}$ in the LG case.
In the Ising case, the estimated errors are slightly larger than 
the discrepancies
between the values obtained with the two measures.
In the LG case, the estimated errors are slightly smaller.
However, larger uncertainties in the error estimates
 appeared in the LG case.
If we use the largest estimates for the errors, our final result 
for the first method is:
\begin{eqnarray}
\gamma &=& 1.299140730159\pm 3\times 10^{-12}\\
\Delta &=& 0.4259485\pm 2\times 10^{-6} \ .
\end{eqnarray}

\section{The Eigenvalues of the Linearized RG Transformation}
\label{sec:eig}

As explained above, the easiest way to calculate the 
critical exponents consists in 
linearizing the RG
transformation near a fixed point $R^{\star}(k)$ specified by 
the coefficients ${a^\star}_ l$. This can be done as follows. First
we express the coefficients after $n$ iterations in terms of
small variations about the fixed point:
\begin{equation}
a_{n,l}=a^\star_l+\delta a_{n,l}\ .
\end{equation}
At the next iteration, we obtain the linear variations
\begin{equation}
\delta a_{n+1,l} = \sum_{m=1}^{l_{max}} M_{l,m} \delta a_{n,m} \ .
\end{equation}
The $l_{max}\times l_{max}$ matrix appearing in this equation is
\begin{equation}
M_{l,m} = {\partial a_{n+1,l}\over \partial a_{n,m} } \ ,
\label{eq:mat}
\end{equation}
evaluated at the  fixed point. 

Approximate fixed points can be found by approaching
$\beta_c$ from below and iterating until the ratio $a_{n+1,1}/a_{n,1}$ 
takes a value which is
as close as possible to 1. This procedure is described in Ref.\cite{rapid}. 
The approximated fixed points obtained with this procedure
depend on $\beta_c$. Using their explicit form 
which we denote $R^{\star}(k,\beta_c)$, we obtained 
a universal 
function $U(k)$
by absorbing $\beta$ into $k$. More explicitly, we found that 
\begin{equation}
U(k)=R^{\star}(\sqrt{\beta_c}k,\beta_c)\ ,
\label{eq:ufonc}
\end{equation}
is in very good approximation independent of the model considered.
This function is related to a fixed point $f(s^2)$ constructed 
in Ref.\cite{wittwer}
by the relation
\begin{equation}
U(k)\propto f(({{c-4}\over{2c}})k^2)  \ .
\label{eq:trans}
\end{equation}
The Taylor coefficients of $f$ can be found in the file \verb+approx.t+
in\cite{wittwer}. Normalizing Eq.(\ref{eq:trans}) 
with $U(0)=1$, we obtain
\begin{equation}
U(k)=1. - 0.35871134988 k^2 + 0.0535372882 k^4 -\dots \ .
\label{eq:ufp}
\end{equation} 

It is not known if there is only one non-trivial fixed point for 
Dyson's model.
Using the parametrization
of Eq. (\ref{eq:lg}), we have considered\cite{rapid} the 12 cases obtained
by choosing among the following possibilities: $m^2=\pm 1 
$ (single or double-well potentials), $p=2,3$ or 4 (coupling constants
of positive, zero and negative dimensions when the cut-off is  restored)
and $g=10$ or 0.1 (moderately large and small couplings).
All approximate fixed points we have constructed give 
a function $U(k)$ very 
close to Eq. (\ref{eq:ufp}). The closeness can be characterized by 
the $\rho$-norms introduced in 
\cite{wittwer}. For $\rho =2$
and $l\leq 42$ we found that the error 
$\delta u_l$ on the $l$-th coefficients of the 
approximate $U(k)$ with respect to the accurate 
expression obtained from Ref.\cite{wittwer} were bounded by 
$|\delta u_l|<{{5\times 10^{-5}}\over {l!2^l}}$ for calculations using double
precision.
In other words, the function $U(k)$ 
seems to be independent of the
general shape of the potential, the strength of the interactions
and whether or not the model is perturbatively renormalizable.

Using these approximate fixed points, we were able to obtain
$\gamma$ and $\Delta$ with 7 decimal points. In the following, we 
will use directly the more precise function $U(k)$ constructed by 
Koch and Wittwer\cite{wittwer}. We retained 16 significant digits
for the coefficients appearing in Eq. (\ref{eq:ufp}) and used values 
of $l_{max}$ up to 65. We then calculated the eigenvalues of the 
matrix given in Eq. (\ref{eq:mat}) with two different methods.
The first was using ``blindly'' the instruction $Eigenvalues$ 
in $Mathematica$. The second consisted in using the eigenvalue routine
$LAPACK$\cite{lapack} for which we were able to vary the control parameters
of the program. The two methods gave identical results with
14 decimal points for the first two eigenvalues. 
The first six eigenvalues are given below.

\begin{quasitable}
\begin{tabular}{lcc}
$n$&$\lambda_n$ \\ \tableline
1&1.42717247817759\\  
2&0.859411649182006\\
3&0.479637305387532\\
4&0.255127961414034\\ 
5&0.131035246260843\\ 
6&0.0654884931298533\\
\end{tabular}
\end{quasitable}

\begin{figure}
\vskip15pt
\centerline{\psfig{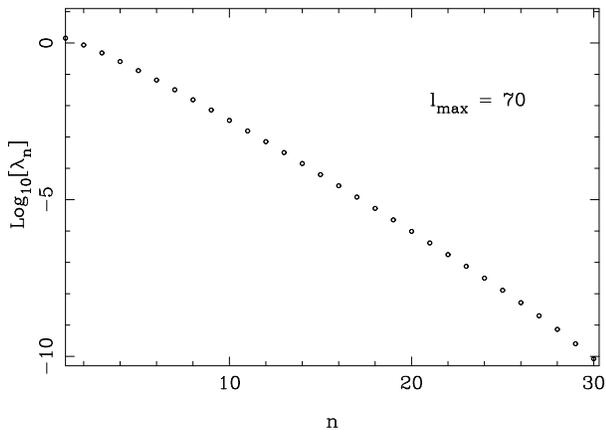}}
\vskip15pt
\caption{log$_{10}[\lambda_n]$
  versus $n$.
} \label{fig:eigen}
\end{figure}

In order to get an idea regarding the 
asymptotic behavior of the eigenvalues, a 
larger set of values is displayed in Fig. \ref{fig:eigen}.

It is clear from the figure that $\lambda_n$ falls faster than 
exponentially with $n$. This property is important when the 
non-linear effects are calculated.

Using the first two eigenvalues and the relationship between the eigenvalues
and the exponents reviewed in the previous section, we obtain the values
\begin{eqnarray}
\gamma &=& 1.2991407301586\pm {10^{-13}}\\
\Delta &=& 0.4259468589881\pm {10^{-13}} \ .
\end{eqnarray}
The (conservative) estimation of the errors is based on  
errors of order $10^{-14}$ on the eigenvalues and the fact that 
the derivatives of the exponents with respect to the eigenvalues yields
factors less than 4.

\section{Conclusions and Perspectives}

In conclusion, 
we have calculated the exponents $\gamma$ and $\Delta$ with a 
accuracy significantly better than in Ref. \cite{rapid}.
The three independent calculations performed here 
agree on the following value for the leading exponent:
\begin{eqnarray}
\gamma &=& 1.299140730159 \pm {10^{-12}} \ .
\end{eqnarray}
Our results show the excellent agreement between the methods developed 
in Ref. \cite{finite} and an expansion about the fixed point of 
Ref. \cite{wittwer}. As far as the calculation of the 
exponents are concerned, the linearization procedure
is much simpler and more accurate. 

It is important to know if 
the non-universal quantities $A_0,\ A_1,\ A_2,\ A_a,\dots$ could also
be calculated by a using an expansion about the fixed point which
involves non-linear terms. We have addressed this question in 
a simplified model, namely the recursion relation for the susceptibility
\begin{equation}
\chi_{n+1}=\chi_n +\beta ({c\over2})^{n+1}\chi_n^2
\end{equation}
A detailed analysis\cite{steve}, shows that in this model, the unknown
quantities appearing in the scaling law for $\chi$ are completely
calculable. If the procedure can be extended to the hierarchical
model, then we could almost consider the model as solvable.

Assuming that the non-universal quantities can be calculated in a
reasonably simple way for all the renormalized quantities, we 
would be in position to decide if the introduction of the bare
parameters can be replaced by a choice of non-universal quantities
appearing in the scaling laws. If we knew the
range of these non-universal quantities 
and their mutual dependence, we could just input an ``independent set'' and 
obtain directly all the scaling laws. 
In particular, in $D=4$, this procedure would yield triviality bounds.
This alternate way of using input parameters in field theory is 
now being investigated.

\acknowledgments
This research was supported in part by the Department of Energy
under Contract No. FG02-91ER40664.
J.J. Godina is supported by
a fellowship from CONACYT. 
Y.M. thanks P. Wittwer and the CERN lattice group 
for useful conversations, the CERN theory division for its 
hospitality while this work was initiated, and the Aspen Center
for Physics, where this manuscript was in part written, for
its stimulating environment. M.B.O. thanks the CERN theory division for its 
hospitality while this work was initiated.

\end{multicols} 
\end{document}